\documentclass[preprint]{elsarticle_new}
\usepackage{amssymb}
\usepackage[final]{epsfig}
\usepackage{color}
\usepackage{graphicx}
\usepackage{lineno}
\newcommand{\beq}{\begin{equation}}
\newcommand{\eeq}{\end{equation}}
\newcommand{\bdi}{\begin{displaymath}}
\newcommand{\edi}{\end{displaymath}}
\newcommand{\no}{\nonumber}
\newcommand{\bea}{\begin{eqnarray}}
\newcommand{\eea}{\end{eqnarray}}
\newcommand{\vep}{\varepsilon}

\newcommand{\text}{}

\newcommand{\de}{\partial}

\begin{document}

\begin{frontmatter}

\title{Green's functions and weighting fields for dielectric multilayer geometries}

\author{W. Riegler}

\address{CERN, Geneva, Switzerland}

\begin{abstract}

The electrostatic potential of a point charge in a geometry with an arbitrary number of parallel layers of different permittivity, bounded by two grounded planes, is derived. Different boundary conditions in the transverse direction are discussed. For the case where a grounded plane is segmented into strips and pads, the solution is used to derive the weighting fields for these electrodes.

\end{abstract}

\end{frontmatter}

\section{Introduction}

\begin{figure}[ht]
 \begin{center}
  \includegraphics[width=6cm]{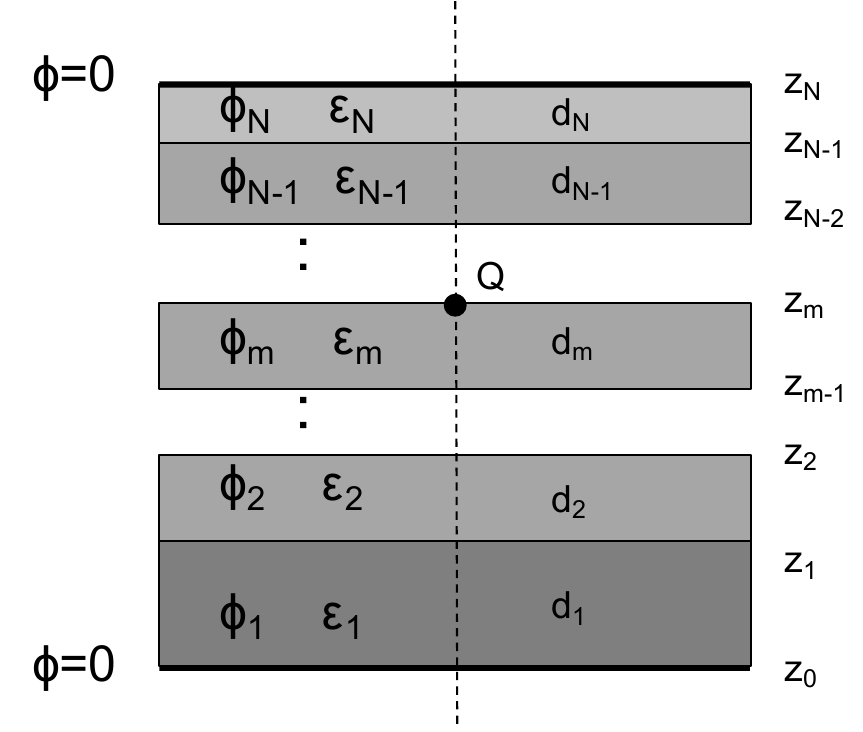}
  \caption{$N$ insulating layers of different dielectric permittivity $\vep_n$ with grounded plates at $z = z_0$ and $z=z_N$. A point charge $Q$ is placed at the boundary between layer $m$ and $m+1$ at $z=z_m$. To describe a point charge inside a dielectric layer, one simply uses the same permittivity in layers $m$ and $m+1$. }
  \label{pointcharge_multilayer}
  \end{center}
\end{figure}

Fig. \ref{pointcharge_multilayer} shows a geometry with $N$ parallel layers having different permittivities $\vep_n$. A point charge $Q$ is placed on the boundary between layer $m$ and layer $m+1$. The potential in each of the $n$ layers is formulated by means of a characteristic function \cite{werner3} 
\beq \label{f_definition} 
         f_{mn}(k, z) = A_{n}(k)e^{kz}+ B_{n}(k)e^{-kz} \qquad n=1,\ldots,N
\eeq
The $2N$ coefficients $A_n(k), B_n(k)$ are defined by a set of $2N$ linear equations. 
This paper presents the solution of these equations and derives explicit expressions for $f_{mn}(k, z)$ for arbitrary $N$, which were not given in Ref. \cite{werner3}. Solutions for three different boundary conditions are then presented: a geometry that is infinitely extended in the $x$- and $y$-directions, a geometry bounded by a grounded cylinder at $r=r_c$, and a geometry bounded by grounded planes at $x=0,a$ and $y=0,b$.
Using the quasi-static approximation of Maxwell's equations, these solutions also apply to the case where the individual layers show weak conductivity. They are therefore well suited for the calculation of electric fields, weighting fields and signals in Resistive Plate Chambers (RPCs), resistive MICROMEGAs and similar detector structures that use resistive electrodes for high-voltage applications, discharge protection or charge spreading.

\section{General Solution}

\noindent
As shown in \cite{werner3}, the equations defining the $2N$ coefficients $A_n(k)$ and $B_n(k)$ are 

\bea \label{defining_equations}
  f_{m1}(k, z_0)  =  0 &&
   f_{mN}(k, z_N)  =  0 \\
   f_{mn}(k,z_n) & = &   f_{m(n+1)}(k,z_n) \\
   \vep_n \frac{\de f_{mn}(k, z)}{\de z }\vert_{z=z_n} & = & \vep_{n+1}  \frac{\de  f_{m(n+1)}(k, z) }{\de z}\vert_{z=z_n}+k\delta_{mn}
\eea
and inserting $f_{mn}(k, z)$ from Eq. \ref{f_definition} gives 
\bea
    B_1  & = &  -A_1 e^{2kz_0}  \label{basic1} \\
   B_N   & =  & -A_N e^{2kz_N}  \label{basic2}  \\
    \phantom{ \vep_n (}A_n e^{kz_n} + B_n e^{-kz_n}  &=& \phantom{\vep_{n+1}(}A_{n+1} e^{kz_n} + B_{n+1} e^{-kz_n}  \label{basic3} \\
   \vep_n (A_ne^{kz_n} -  B_n e^{-kz_n} ) &=& \vep_{n+1}(A_{n+1} e^{kz_n} -  B_{n+1}
  e^{-kz_n}) + \delta_{nm} \label{basic4}
\eea
for $n=1...N-1$. The charge $Q$ is located on the boundary between layer $m$ and $m+1$. The $\vep_n$ refer to the dimensionless relative permittivity, so the $f_{mn}(k, z)$ are dimensionless. We can use these expressions to relate $A_{n+1}, B_{n+1}$ to $A_n, B_n$ for interfaces other than $m$ ($n \neq m \rightarrow \delta_{mn}=0$).
\beq
       \left(
   \begin{array}{c}
       A_{n+1} \\
       B_{n+1}
   \end{array}
   \right) = \frac{1}{2 \vep_{n+1}}\left(
   \begin{array}{cc}
        \vep_{n+1}+\vep_{n} &  e^{-2kz_n} ( \vep_{n+1}-\vep_{n} ) \\[5mm]
   e^{2kz_n} ( \vep_{n+1}-\vep_{n} )    &   \vep_{n+1}+\vep_{n} 
   \end{array}\right)
     \left( \begin{array}{c}
       A_{n} \\
       B_{n}
   \end{array}
   \right)
\eeq
The inverse relation, which expresses $A_n,B_n$ in terms of $A_{n+1},B_{n+1}$, is 
\beq
       \left(
   \begin{array}{c}
       A_{n} \\[5mm]
       B_{n}
   \end{array}
   \right) = \frac{1}{2 \vep_{n}}\left(
   \begin{array}{cc}
        \vep_{n+1}+\vep_{n} &  e^{-2kz_n} ( \vep_{n}-\vep_{n+1} ) \\[5mm]
   e^{2kz_n} ( \vep_{n}-\vep_{n+1} )    &   \vep_{n+1}+\vep_{n} 
   \end{array}\right)
     \left( \begin{array}{c}
       A_{n+1} \\[5mm]
       B_{n+1}
   \end{array}
   \right)
\eeq
By iterating this matrix multiplication, we can relate the coefficients $A_n, B_n$ to $A_1, B_1$ according to 
\beq \label{up}
   A_n = u_A(n) A_1 \qquad B_n = u_B(n) A_1 \qquad 1<n \le m
\eeq
and we can relate the coefficients $A_n, B_n$ to $A_N, B_N$ according to 
\beq \label{down}
   A_n = d_A(n) A_N \qquad B_n = d_B(n) A_N \qquad m+1 \le n<N
\eeq
The expressions $u_A, u_B, d_A, d_B$ can be calculated iteratively and are given by
\beq
     u_A(n) =   \sum_{i=1}^{2^{n-1} } c_{ni}e^{kv_{ni}}    \qquad u_B(n) =  - \sum_{i=1}^{2^{n-1} } c_{ni}e^{-k(v_{ni}-2z_0)} 
\eeq
\beq 
  d_A(n) =   \sum_{i=1}^{2^{N-n} } g_{ni}^{(N)}e^{-kw_{ni}^{(N)}}  \qquad
  d_B(n)  =  -\sum_{i=1}^{2^{N-n} } g_{ni}^{(N)}e^{k(w_{ni}^{(N)}+2z_N)}   
\eeq
where the four sets of coefficients $c_{ni}, v_{ni}, g_{ni}^{(N)}, w_{ni}^{(N)}$ are defined by
\beq \label{matr1}
  c_{ni} = \frac{1}{2^{n-1}} \prod_{j=1}^{n-1} \frac{(\vep_j+\sigma_{ij}^{(n)} \vep_{j+1})}{\vep_{j+1}}
  \qquad v_{ni} = \sum_{j=1}^{n-1} (\theta_{ij}^{(n)}-1)d_j
\eeq
\beq \label{matr2}
  g_{ni}^{(N)} = \frac{1}{2^{N-n}} \prod_{j=1}^{N-n} \frac{(\vep_{N+1-j}+\sigma_{ij}^{(N+1-n)} \vep_{N-j})}{\vep_{N-j}}
  \qquad w_{ni}^{(N)} = \sum_{j=1}^{N-n} (\theta_{ij}^{(N+1-n)}-1)d_{N+1-j}
\eeq
The matrix $\sigma^{(n)}_{ij}$ is defined as a matrix with $n-1$ columns and $2^{n-1}$ rows containing all combinations of $1$ and $-1$, which can be derived from the binary representation of all numbers from $0$ to $2^{n-1}-1$. 
As examples, for $n=2,3$ we have
\beq
    \sigma^{(2)}_{ij} = 
   1-2\times \left(
     \begin{array}{c}
     0 \\ 1
     \end{array}
   \right) =     \left(
     \begin{array}{c}
     1 \\ -1
     \end{array}
    \right) 
    \quad
          \sigma^{(3)}_{ij} = 
    1-2\times\left(
     \begin{array}{cc}
     0 & 0 \\  0 & 1 \\ 1 & 0 \\ 1 & 1
     \end{array}
    \right) =    \left(
     \begin{array}{cc}
    1 & 1 \\  1 & -1 \\ -1 & 1 \\ -1 & -1
     \end{array}
    \right) 
\eeq
The matrix $\theta_{ij}^{(n)}$ is defined by 
\beq
   \theta^{(n)}_{ij} = \prod_{l=j}^{n-1} \sigma^{(n)}_{il} 
\eeq
For $n=2, 3$ we have
\beq
        \theta^{(2)}_{ij} = 
    \left(
     \begin{array}{c}
     1 \\ -1
     \end{array}
    \right) 
   \quad
         \theta^{(3)}_{ij} = 
    \left(
     \begin{array}{cc}
     1 & 1 \\  -1 & -1 \\ -1 & 1 \\ 1 & -1
     \end{array}
    \right) 
\eeq
At the interface carrying the charge, $z=z_m$, Eq. \ref{basic4} reads as 
\bea \label{interface_equation} 
    A_me^{k z_m} + B_m e^{-k z_m} & = &   A_{m+1}e^{k z_m} + B_{m+1} e^{-k z_m} \\
      \vep_m(A_me^{k z_m} - B_m e^{-k z_m} )& = &   \vep_{m+1}(A_{m+1}e^{k z_m} - B_{m+1} e^{-k z_m} ) + 1 \no
\eea
We now use the relations from Eqs. \ref{up} and \ref{down} 
\beq
   A_m=u_A(m) A_1 \qquad   B_m=u_B(m) A_1 
\eeq
\beq
   A_{m+1}=d_A(m+1) A_N  \qquad   B_{m+1}=d_B(m+1) A_N 
\eeq
These relations determine $A_1$ and $A_N$ from Eq. \ref{interface_equation}, and consequently all coefficients $A_n$ and $B_n$. Inserting these coefficients into Eq. \ref{f_definition} yields the final expressions for $f_{mn}(k, z)$:
\\ \\ 
For $ 1 \le n \le m$ 
\beq \label{final1}
         f_{mn}(k, z)= \frac{
         \left(\sum_{i=1}^{2^{N-m-1}} g_{(m+1)i}^{(N)} \sinh [k( w_{(m+1)i}^{(N)}+z_N-z_m)] \right)
         \left( \sum_{i=1}^{2^{n-1}} c_{ni}\sinh[k( v_{ni}+z-z_0)] \right)
         }{
         \vep_N\sum_{i=1}^{2^{N-1}}c_{Ni}\sinh [k \, ( v_{Ni}+z_N-z_0)]
         }
\eeq
and for $m+1\le n \le N$
\beq \label{final2}
         f_{mn}(k, z)=\frac{ 
         \left( \sum_{i=1}^{2^{N-n}}  g_{ni}^{(N)}\sinh[k( w_{ni}^{(N)}+z_N-z)] \right)
         \left( \sum_{i=1}^{2^{m-1}} c_{mi} \sinh [k( v_{mi}+z_m-z_0)] \right) 
         }{
         \vep_N\sum_{i=1}^{2^{N-1}}c_{Ni}\sinh [k \, ( v_{Ni}+z_N-z_0)]
         }
\eeq
with $z_n = z_0 + \sum_{j=1}^n d_j$. Some explicit examples of the coefficients $c_{ni}, v_{ni}, g_{ni}^{(N)}, w_{ni}^{(N)}$ are given in the Appendix.
When evaluating these expressions numerically, one can verify their correctness by checking whether the numerical results satisfy Eq. \ref{defining_equations}.

\section{Behaviour of $f_{mn}$ for small and large values of $k$} 

In the expressions for the potentials and fields, $f_{mn}(k,z)$ is evaluated over the range $0{<}k{<}\infty$. It is therefore important for convergence to understand its behaviour for both small and large $k$. Since $\sinh(x) \approx x$ for small values of $x$, the functions $f_{mn}(k,z)$ are proportional to $k$ and therefore
\beq
      \lim_{k \rightarrow 0} f_{mn}(k, z) =0
\eeq
For large values of $k$ the numerator and the denominator diverge, and we have to find the largest arguments of the $\sinh(x)$ functions and approximate them by $\sinh(x) \approx e^x/2$. Since the largest exponents occur for $i=1$, for which $v_{mi} = w_{(m+1)i}^{(N)}=0$, we find the following approximate expressions for large values of $k$: 
\\ \\
For $1 \le n \le m$
\beq
       f_{mn} (k, z) \approx \frac{1}{2} \frac{g_{(m+1)1}^{(N)} e^{k(z_N-z_m)}\, c_{n1}e^{k(z-z_0)}}{\vep_N\,c_{N1}e^{k(z_N-z_0)}} 
   = \frac{1}{2} \frac{g_{(m+1)1}^{(N)}  c_{n1}}{\vep_N\,c_{N1}}\,e^{-k(z_m-z)}
\eeq
For $m+1 \le n \le N$
\beq
  f_{mn} (k, z) \approx \frac{1}{2} \frac{c_{m1}e^{k(z_m-z_0)} g_{n1}^{(N)} e^{k(z_N-z)}}{\vep_N\,c_{N1}e^{k(z_N-z_0)}} 
   = \frac{1}{2} \frac{c_{m1}  g_{n1}^{(N)}}{\vep_N \, c_{N1}}\,e^{-k(z-z_m)}
\eeq
We learn that for large values of $k$ the functions decay exponentially as $e^{-k\vert z-z_m \vert}$, so the integrals and sums have to extend to values of $k$ for which $k\vert z-z_m\vert$ is at least a few units. 
\\  
For $z=z_m$, corresponding to layers $n=m$ and $n=m+1$, the asymptotic expression is independent of $k$ and therefore does not approach zero as $k\rightarrow\infty$, resulting in poor convergence of the corresponding integrals. For these values of $z$, the functions are approximated by  
\\ \\
\bea \label{approx1}
   f_{mm} (k, z) &\approx & \frac{1}{2} \frac{g_{(m+1)1}^{(N)}  c_{m1}}{\vep_N c_{N1}}\,e^{-k(z_m-z)} = \frac{e^{-k(z_m-z)} }{\vep_m+\vep_{m+1}}
   \label{approx2} \\
     f_{m(m+1)} (k, z) &\approx  &
   \frac{1}{2} \frac{c_{m1}  g_{(m+1)1}^{(N)}}{\vep_N c_{N1}}\,e^{-k(z-z_m)}=\frac{e^{-k(z-z_m)} }{\vep_m+\vep_{m+1}}
\eea
We will use these expressions in the following.

\section{General solution} 

\begin{figure}[ht]
 \begin{center}
    \includegraphics[width=10cm]{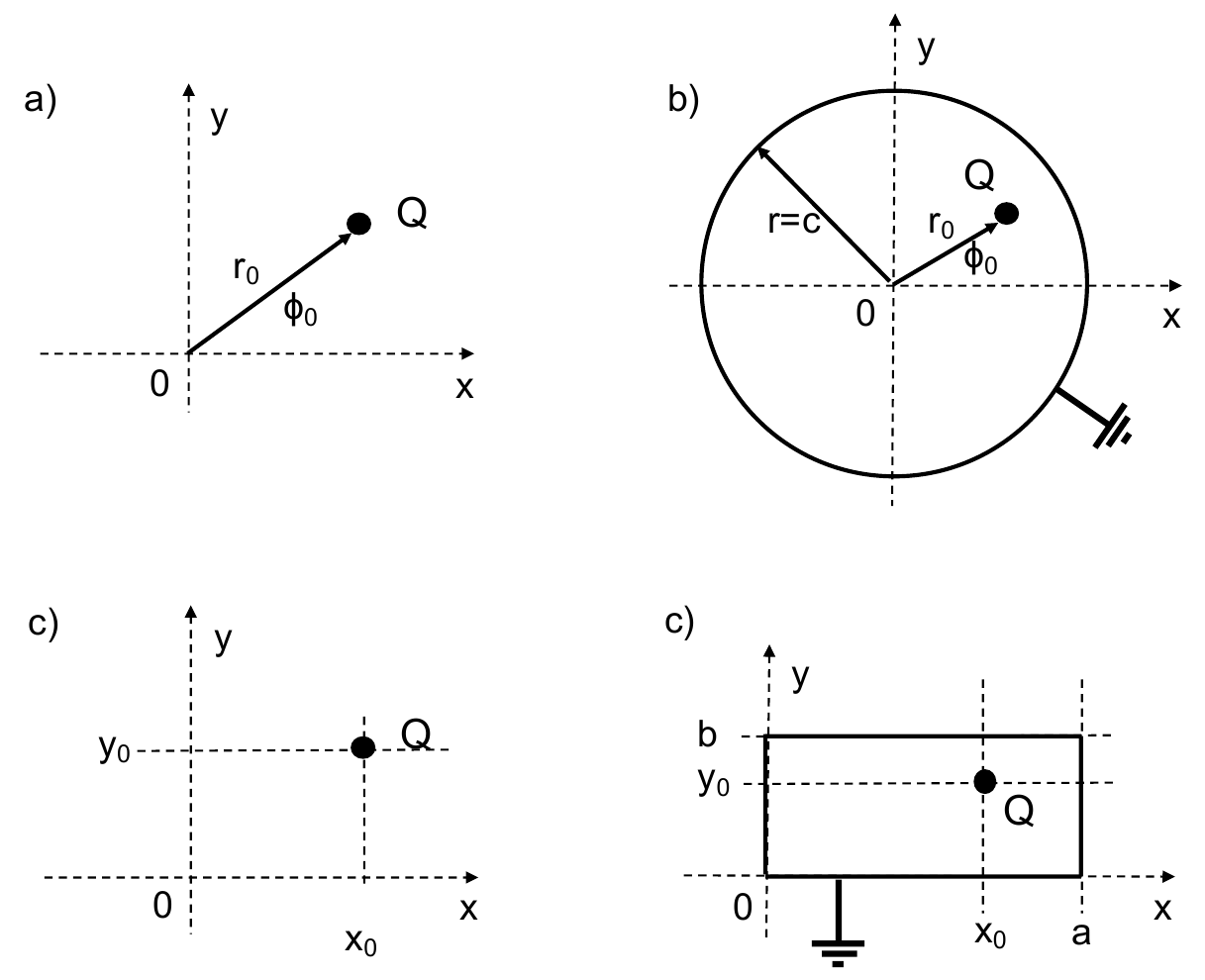}
  \caption{Coordinate definitions for different boundary conditions of the multilayer geometry in the $x, y$ directions. a) Infinitely extended geometry in cylindrical coordinates b) Geometry grounded on a cylinder at $r{=}r_c$ in cylindrical coordinates c) Infinitely extended geometry in Cartesian coordinates d) Geometry grounded at $x{=}0, a$, $y{=}0, b$ in Cartesian coordinates.}
  \label{xy_plane_definitions}
  \end{center}
\end{figure}

We can now write the potential for the geometry shown in Fig. \ref{pointcharge_multilayer} as 
\beq
   \phi_{mn}(x, y, z)=\frac{Q}{\vep_0}\,G_{mn}(x, y, z) \qquad z_{n-1}<z<z_n
\eeq
where $G_{mn}$ is expressed in terms of the function $f_{mn}(k, z)$. In the following we discuss four different boundary conditions, as shown  in Fig. \ref{xy_plane_definitions}.

\subsection{Infinitely extended geometry in cylindrical coordinates}

If the geometry in Fig. \ref{pointcharge_multilayer} is infinitely extended in the $x$- and $y$-directions, the Green's function at the position $(r,\varphi,z)$ for a point charge at $(r_0,\varphi_0,z_m)$ is given by (cylindrical coordinates):
\beq \label{sol0}
    G_{mn}(r, r_0, \varphi, \varphi_0, z, z_m) = \frac{1}{2 \pi } \int_0^\infty \sum_{\alpha=-\infty}^\infty
    e^{i\alpha (\varphi-\varphi_0)} J_\alpha(kr)J_\alpha(kr_0)f_{mn}(k,z) dk 
\eeq
If the charge is placed at $r_0=0$, we have 
\beq \label{sol1}
    G_{mn}(r,  z, z_m) = \frac{1}{2 \pi }\int_0^\infty J_0(kr)f_{mn}(k,z) dk 
\eeq
In order to improve the convergence of the integral for large values of $k$ we use the method introduced in Ref. \cite{schnizer1}, based on the following relation \cite{watson}:
\beq
   \frac{1}{\sqrt{r^2+r_0^2-2rr_0\cos (\varphi-\varphi_0)+(z-z_m)^2}} =  \int_0^\infty \sum_{\alpha=-\infty}^\infty
    e^{i\alpha (\varphi-\varphi_0)} J_\alpha(kr)J_\alpha(kr_0)e^{-k\vert z-z_m\vert} dk 
\eeq
We can therefore rewrite Eq. \ref{sol0} by adding and subtracting Eq. $\ref{approx1}$ in the integrand and writing
\\ \\
For $1 \le n \le m$:
\bea
    G_{mn}(r, r_0, \varphi, \varphi_0, z, z_m)  & = &   
    \frac{1}{2\pi} \frac{g_{(m+1)1}^{(N)}  c_{n1}}{2\vep_N\,c_{N1}} \frac{1}{ \sqrt{r^2+r_0^2-2rr_0\cos (\varphi-\varphi_0)+(z-z_m)^2}} \\
     &+& \frac{1}{2 \pi } \int_0^\infty \sum_{\alpha=-\infty}^\infty
    e^{i\alpha (\varphi-\varphi_0)} J_\alpha(kr)J_\alpha(kr_0) \left( f_{mn}(k,z)  -  \frac{1}{2} \frac{g_{(m+1)1}^{(N)}  c_{n1}}{\vep_N\,c_{N1}}\,e^{-k(z_m-z)} \right)dk \nonumber
\eea
\\ \\
For $m+1 \le n \le N$:
\bea
    G_{mn}(r, r_0, \varphi, \varphi_0, z, z_m)  & = &   
    \frac{1}{2\pi} \frac{c_{m1}  g_{n1}^{(N)}}{2\vep_N \, c_{N1}}\frac{1}{ \sqrt{r^2+r_0^2-2rr_0\cos (\varphi-\varphi_0)+(z-z_m)^2}} \\
     &+& \frac{1}{2 \pi } \int_0^\infty \sum_{\alpha=-\infty}^\infty
    e^{i\alpha (\varphi-\varphi_0)} J_\alpha(kr)J_\alpha(kr_0) \left( f_{mn}(k,z)  -   \frac{1}{2} \frac{c_{m1}  g_{n1}^{(N)}}{\vep_N \, c_{N1}}\,e^{-k(z-z_m)} \right)dk \nonumber
\eea
These expressions show much improved convergence for large values of $k$, especially for small values of $\vert z-z_m \vert$.

\subsection{Geometry grounded on a cylinder at $r=r_c$ in cylindrical coordinates}  

If the geometry from Fig. \ref{pointcharge_multilayer} is grounded at $r=r_c$ and the point charge $Q$ is placed at $r_0, \varphi_0, z_m)$, we have 
\beq
   G_{mn}(r, r_0, \varphi, \varphi_0, z, z_m) = \frac{1}{\pi r_c  } \sum_{l=1}^\infty \sum_{\alpha=-\infty}^\infty
    e^{i\alpha (\varphi-\varphi_0)}  \frac{J_\alpha(k_{\alpha l}r)J_\alpha(k_{\alpha l}r_0)}{j_{\alpha l}[J_{\alpha +1}(j_{\alpha l})]^2} f_{mn}(k_{\alpha l},z)  
\eeq
where $ k_{\alpha l}=\frac{j_{\alpha l}}{r_c} $ with $j_{\alpha l}$ being the $l^{\mathrm{th}}$ zero of $J_\alpha(x)$. If the charge is located at $r_0=0$ we have
\beq
   G_{mn}(r, z, z_m) = \frac{1}{\pi r_c  } \sum_{l=1}^\infty
     \frac{J_0(k_{0 l}r)}{j_{0l}[J_{1}(j_{0 l})]^2} f_{mn}(k_{0 l},z)  
\eeq
For small values of $\vert z - z_m \vert$ the series shows rather slow convergence since the expression behaves as $1/l^{3/2}$.

\subsection{Infinitely extended geometry in Cartesian coordinates}

For the case where the geometry of Fig. \ref{pointcharge_multilayer} is infinitely extended and we use Cartesian coordinates for a point charge located at $(x_0,y_0,z_m)$, we have 
\beq
    G_{mn}(x, x_0, y, y_0, z, z_m) = \frac{1}{\pi^2 } \int_0^\infty \int_0^\infty
    \cos[k_x(x-x_0)]  \cos[k_y(y-y_0)]
    \frac{f_{mn}(k, z)}{k} dk_x dk_y 
\eeq
with $k=\sqrt{k_x^2+k_y^2}$. We can again rewrite the expressions in a form that shows better convergence for large values of $k$ as 
\\ \\
For $1 \le n \le m$:
\bea
    G_{mn}(x, x_0, y, y_0, z, z_m)  & = &   
    \frac{1}{2\pi} \frac{g_{(m+1)1}^{(N)}  c_{n1}}{2\vep_N\,c_{N1}} \frac{1}{ \sqrt{(x-x_0)^2+(y-y_0)^2+(z-z_m)^2}} \\
     &+& \frac{1}{\pi^2 } \int_0^\infty \int_0^\infty \cos[k_x(x-x_0)]  \cos[k_y(y-y_0)]  \\
     &\times&\frac{1}{k}
  \left( f_{mn}(k,z)  -  \frac{1}{2} \frac{g_{(m+1)1}^{(N)}  c_{n1}}{\vep_N\,c_{N1}}\,e^{-k(z_m-z)} \right)dk_xdk_y \nonumber
\eea
\\ \\
For $m+1 \le n \le N$:
\bea
    G_{mn}(x, x_0, y, y_0, z, z_m)  & = &   
    \frac{1}{2\pi} \frac{c_{m1}  g_{n1}^{(N)}}{2\vep_N \, c_{N1}}\frac{1}{ \sqrt{(x-x_0)^2+(y-y_0)^2+(z-z_m)^2}} \\
     &+& \frac{1}{\pi^2 } \int_0^\infty \int_0^\infty \cos[k_x(x-x_0)]  \cos[k_y(y-y_0)] \\ 
     &\times & \frac{1}{k}\left( f_{mn}(k,z)  -   \frac{1}{2} \frac{c_{m1}  g_{n1}^{(N)}}{\vep_N \, c_{N1}}\,e^{-k(z-z_m)} \right)dk_xdk_y \nonumber
\eea

\subsection{Geometry grounded at  $x=0, a$ and $y=0, b$ in Cartesian coordinates}

Assuming that the geometry of Fig. \ref{pointcharge_multilayer} is grounded at $x=0,a$ and $y=0,b$ and that we have a point charge at $(x_0,y_0,z_m)$, we find
\beq  \label{sol4}
    G_{mn}(x, x_0, y, y_0, z, z_m) = \frac{4}{ab\, } \sum_{\alpha=1}^\infty \sum_{\beta =1}^\infty 
    \sin \left( \alpha \pi \frac{x}{a}\right)
    \sin \left( \alpha \pi \frac{x_0}{a}\right)
    \sin \left( \beta \pi \frac{y}{b}\right)
    \sin \left( \beta \pi \frac{y_0}{b}\right)
    \frac{f_{mn}(k_{\alpha \beta}, z)}{k_{\alpha \beta}} 
    \qquad   \no
\eeq
with $k_{\alpha \beta }=\pi \sqrt{\frac{\alpha^2}{a^2 }+\frac{\beta ^2}{b^2 } }$. For $z=z_m$ we have $f_{mn}$ independent of $k_{\alpha \beta}$ and the terms of the series behave as $1/k_{\alpha \beta}$ resulting in rather slow convergence.

\clearpage

\section{Potential of a thin layer of charge at $z=z_m$}

\begin{figure}[ht]
 \begin{center}
 a)
  \includegraphics[width=6cm]{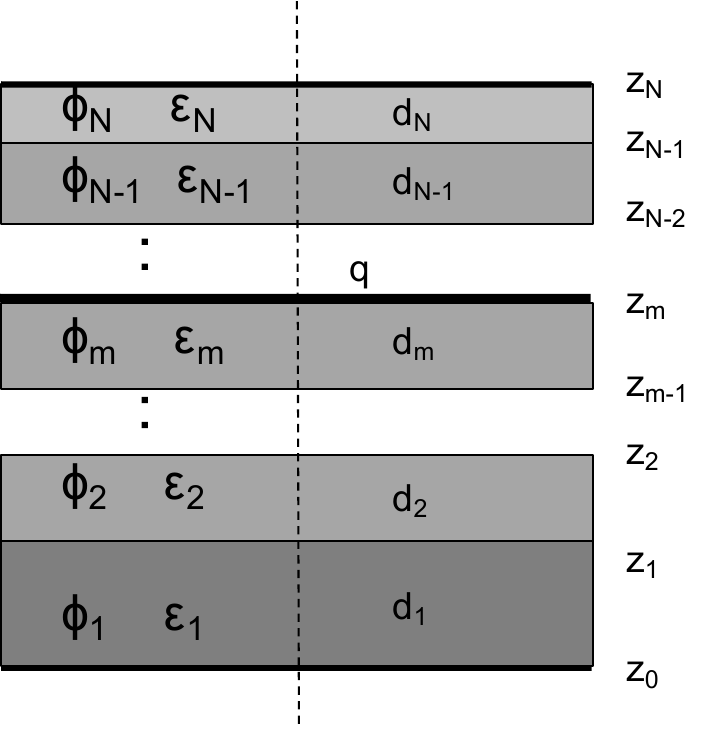}
  \caption{A thin layer of charge with charge density $q$\,[C/cm$^2$] on the boundary between layers $m$ and $m+1$. }
  \label{sheetcharge_multilayer}
  \end{center}
\end{figure}
We now assume an infinitely extended layer of constant surface charge density $q$\,[C/cm$^{2}$] at the interface between layers $m$ and $m+1$ instead of a point charge $Q$. This relation is relevant, e.g., for the calculation of the voltage drop in a detector due to uniform charge deposition. Since the electric field $E_n$ in the different layers is directed only along the $z$-direction, we can use Gauss's law to find the potential $\phi_m(z_m)$ on the surface of this layer. Since the top and bottom plates are grounded we have
\beq
      E_1d_1+E_2d_2+...+E_Nd_N=0
\eeq
and using Gauss's law at the interfaces we have 
\beq
    \vep_{n+1}E_{n+1} =\vep_nE_n\quad n \neq m \qquad   \vep_{m+1}E_{m+1}=\vep_mE_m+\frac{q}{\vep_0}
\eeq
The potential of the layer is then
\beq
    \phi_m =- \sum_{n=1}^m d_nE_n = \sum_{n=m+1}^N d_nE_n
\eeq
The result is then 
\beq
      \phi_m = q/c \qquad c=c_1+c_2=\frac{\vep_0}{\sum_{n=1}^m\frac{d_n}{\vep_n}} + \frac{\vep_0}{\sum_{n=m+1}^N\frac{d_n}{\vep_n}}
\eeq
where $c_1$ and $c_2$ represent the capacitance per unit area between the interface layer (if it were a metal plate) and the bottom and top ground plates, respectively. $c$ represents the total capacitance between the interface layer and ground. On the other hand, we can calculate $\phi_m$ from the above Green's function
\beq
    \phi_m = \frac{q}{2\pi \vep_0} \int_0^{2\pi} \left[\int_0^\infty 2\pi r J_0(kr)dr \right]f_{mm}(k, z_m) dk = 
    \frac{q}{\vep_0} \int_0^\infty \frac{\delta(k)}{k} f_{mm}(k, z_m) dk = \frac{q}{\vep_0} \lim_{k \rightarrow 0} \frac{f_{mm}(k, z_m)}{k}
\eeq
with
\bdi
         f_{mm}(k, z)= \frac{
         \left(\sum_{i=1}^{2^{N-m-1}} g_{(m+1)i}^{(N)} \sinh [k( w_{(m+1)i}^{(N)}+z_N-z_m)] \right)
         \left( \sum_{i=1}^{2^{m-1}} c_{mi}\sinh[k( v_{mi}+z_m-z_0)] \right)
         }
         {
         \vep_N\sum_{i=1}^{2^{N-1}}c_{Ni}\sinh [k \, ( v_{Ni}+z_N-z_0)]
         }
\edi
For small values of $k$ we have
\bea
        f_{mm}(k, z) & \approx &  \frac{
         \left(\sum_{i=1}^{2^{N-m-1}} g_{(m+1)i}^{(N)} k( w_{(m+1)i}^{(N)}+z_N-z_m)\right)
         \left( \sum_{i=1}^{2^{m-1}} c_{mi}k( v_{mi}+z_m-z_0) \right)
         }
         {
         \vep_N\sum_{i=1}^{2^{N-1}}c_{Ni}k \, ( v_{Ni}+z_N-z_0)
         } \\
         & = & 
         \frac{\left( k\vep_N\sum_{i=m+1}^N \frac{d_i}{\vep_i} \right)\left(k \vep_1\sum_{i=1}^m \frac{d_i}{\vep_i}  \right)  }{\left( k\vep_1\vep_N\sum_{i=1}^N  \frac{d_i}{\vep_i}  \right) }
          = \frac{k \vep_0}{c}
\eea
So we have
\beq
  \phi_m = \frac{q}{\vep_0} \lim_{k \rightarrow 0} \frac{f_{mm}(k, z_m)}{k} = \frac{q}{c}
\eeq
which coincides with the relation derived above.

\section{Charge distributions in the ground planes, weighting fields}

\begin{figure}[ht]
 \begin{center}
 a)
  \includegraphics[width=6cm]{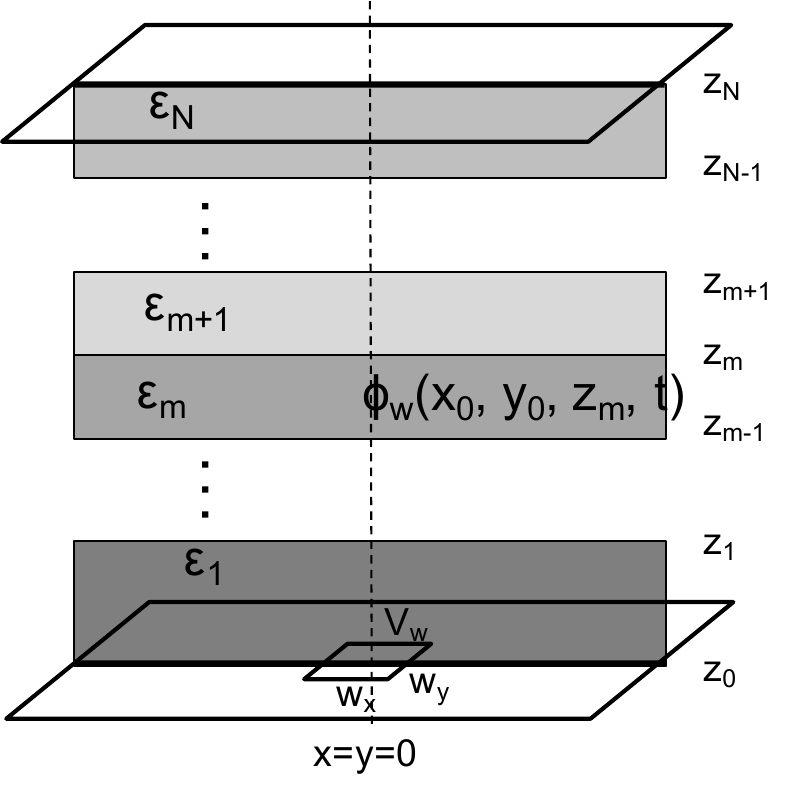}
b)
    \includegraphics[width=6cm]{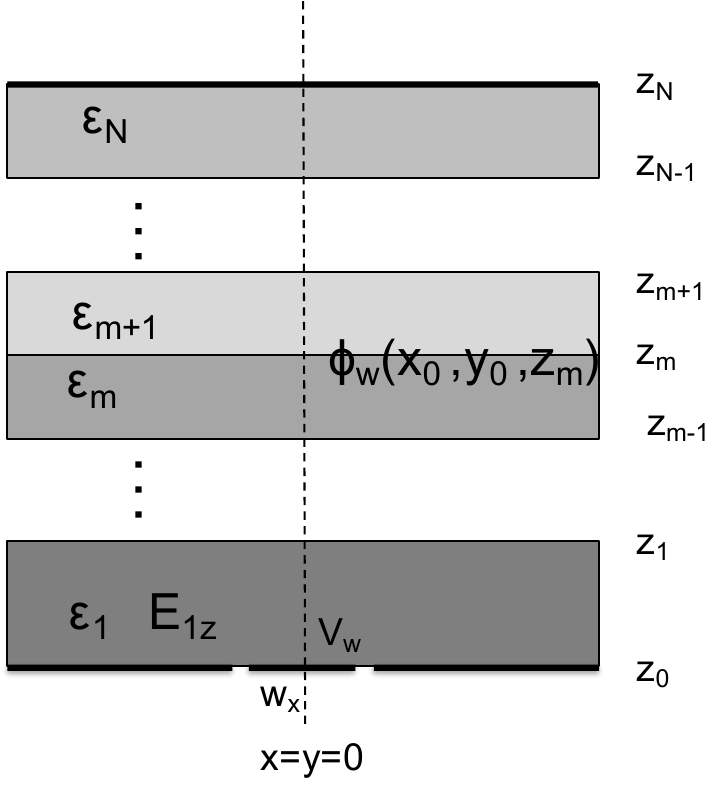}
  \caption{a) A pad of dimensions $w_x$ and $w_y$ centred at the origin; b) a strip of width $w_x$ centred at $x=0$ and infinitely extended in the $y$-direction. }
  \label{weighting_field_pad1}
  \end{center}
\end{figure}

\noindent
By segmenting the bottom electrode into pads or strips as shown in Fig. \ref{weighting_field_pad1}, we can calculate the total induced charge on such an element by integrating the electric field on the surface of the electrode over the area of the electrode:
\beq
    Q^{ind}(x_0, y_0, z_m) = -\vep_0\vep_1 \int_A \frac{\de \phi_{m1}(x, x_0, y, y_0, z, z_m)}{\de z} \vert_{z=z_0} dA
\eeq
By the reciprocity theorem this expression, multiplied by $-V_w/Q$, is equivalent to the potential $\phi^w(x_0, y_0, z_m)$ for the case where the charge is removed and the pad is set to the potential $V_w$. 
\beq
     \phi^w(x_0, y_0, z_m) = -\frac{V_w}{Q} Q^{ind}(x_0, y_0, z_m) = \vep_0\vep_1 \frac{V_w}{Q} \int_A \frac{\de \phi_{m1}(x, x_0, y, y_0, z, z_m)}{\de z} \vert_{z=z_0} dA
\eeq
For the geometry shown in Fig. \ref{weighting_field_pad1}a), where a pad of dimensions $w_x$ and $w_y$ is centred at the origin in the infinitely extended geometry, we have 
\bea
     && \phi^w(x_0, y_0, z_m) =  \vep_1 V_w  \int_{-w_x/2}^{w_x/2}  \int_{-w_y/2}^{w_y/2} \frac{\de G_{m1}(x, x_0, y, y_0, z, z_m)}{\de z}\vert_{z=z_0} dx dy  \\
     & & = \frac{4V_w}{\pi^2 } \int_0^\infty \int_0^\infty
    \cos(k_x x_0)\sin \left(\frac{k_x w_x}{2} \right) 
     \cos(k_y y_0)\sin \left(\frac{k_y w_y}{2} \right) 
    \frac{1}{k_xk_y}\frac{\vep_1}{k}\frac{\de f_{m1}(k, z)}{\de z} \vert_{z=z_0}dk_x dk_y \nonumber
\eea
From Eq. \ref{final1} we have
\beq \label{hm}
 h_m(k) := \frac{\vep_1}{k} \frac{\de f_{m1}(k, z)}{\de z} \vert_{(z=z_0)} =   \frac{
         \vep_1\sum_{i=1}^{2^{N-m-1}} g_{(m+1)i}^{(N)} \sinh [k( w_{(m+1)i}^{(N)}+z_N-z_m)] 
         }{
         \vep_N\sum_{i=1}^{2^{N-1}}c_{Ni}\sinh [k \, ( v_{Ni}+z_N-z_0)]
         }   
\eeq
For a strip that extends to infinity in the $y$-direction, as shown in Fig. \ref{weighting_field_pad1}b), we have
\beq
     \phi^w(x_0,  z_m) =\frac{2V_w}{\pi } \int_0^\infty 
    \cos(k  x_0)\sin \left(\frac{k  w_x}{2} \right) 
    \frac{h_m(k) }{k}dk\nonumber
\eeq
For the geometry that is grounded at $x=0,a$ and $y=0,b$, with a pad of dimensions $w_x$ and $w_y$ centred at $(x_p,y_p)$, using Eq. \ref{sol4}, the weighting potential is
\bea
   &&   \phi^w(x_0, y_0, z_m)  =   \vep_1 V_w  \int_{x_p-w_x/2}^{x_p+w_x/2}  \int_{y_p-w_y/2}^{y_p+w_y/2} \frac{\de G_{m1}(x, x_0, y, y_0, z, z_m)}{\de z}\vert_{z=z_0} dx dy\\
    &&   =  
       \frac{16V_w}{\pi^2}   \sum_{\alpha=1}^\infty \sum_{\beta =1}^\infty 
  \frac{
    \sin \left( \alpha \pi \frac{w_x}{2a}\right)
    \sin \left( \alpha \pi \frac{x_p}{a}\right)
      \sin \left( \alpha \pi \frac{x_0}{a}\right)
      }{\alpha}
      \frac{
    \sin \left( \beta \pi \frac{w_y}{2b}\right)
    \sin \left( \beta \pi \frac{y_p}{b}\right)
     \sin \left( \beta \pi \frac{y_0}{b}\right)
     }{\beta} h_m(k_{\alpha \beta})  \no
\eea

\section{Summary}

\noindent
An analytic expression for the potential of a point charge in a multilayer structure containing an arbitrary number of layers with different permittivities has been derived. Within the quasi-static approximation to Maxwell's equations, the result also describes the time-dependent potential when the layers have finite conductivity. The expressions can be used to calculate electric fields, weighting fields, and signals for a wide range of detectors. Their convergence properties have been discussed, together with methods for improving numerical convergence. The resulting expressions are well suited for numerical evaluation.

\section{Appendix A}
 
 \noindent
 As examples, we give the key expressions from Eqs. \ref{matr1}, \ref{matr2}, \ref{final1}, \ref{final2} and \ref{hm} for geometries with two and three layers. We use $z_n = z_0 + \sum_{j=1}^n d_j$, where $d_j$ is the thickness of layer $j$. \\ \\
For $N=2$ we have
\beq
  c_{ni} = \left(\begin{array}{cc}
    1 & \frac{\vep_1+\vep_2}{2\vep_2} \\
    1 &  \frac{\vep_1-\vep_2}{2\vep_2} 
  \end{array}\right)^T \quad
    v_{ni} = \left(\begin{array}{cc}
    0 & 0 \\
    0 &  -2d_1
  \end{array}\right)^T \quad
  g_{ni} = \left(\begin{array}{cc}
    \frac{\vep_1+\vep_2}{2\vep_1}  & 1 \\
     \frac{-\vep_1+\vep_2}{2\vep_1}  & 1
  \end{array}\right)^T \quad
  w_{ni} = \left(\begin{array}{cc}
    0 & 0\\
    -2d_2 &  0
  \end{array}\right)^T \quad
\eeq
\bea
   D(k) & = &   \frac{(\vep_1-\vep_2)\sinh(k(-d_1+d_2))}{2 \vep_2}+\frac{(\vep_1+\vep_2)\sinh(k(d_1+d_2))}{2\vep_2} \\
   f_{11}(k, z) & = &  \frac{\sinh(kd_2) \sinh(k(z-z_0))}{\vep_2 D(k)} \\
   f_{12}(k, z) & = & \frac{\sinh(kd_1) \sinh(k(z_0+d_1+d_2-z))}{\vep_2 D(k)}   \\
   h_1(k)&=&
{\varepsilon_1\over\varepsilon_2D(k)}
\sinh\left(kd_2\right)
\eea
\\ \\
For  $N=3$ we have
\beq
c_{ni}=\left(
\begin{array}{ccc}
 1 & \frac{\vep_1+\vep_2}{2 \vep_2} & \frac{(\vep_1+\vep_2) (\vep_2+\vep_3)}{4 \vep_2 \vep_3} \\
 1 & \frac{\vep_1-\vep_2}{2 \vep_2} & \frac{(\vep_1+\vep_2) (\vep_2-\vep_3)}{4 \vep_2 \vep_3} \\
 1 & \frac{\vep_1+\vep_2}{2 \vep_2} & \frac{(\vep_1-\vep_2) (\vep_2+\vep_3)}{4 \vep_2 \vep_3} \\
 1 & \frac{\vep_1-\vep_2}{2 \vep_2} & \frac{(\vep_1-\vep_2) (\vep_2-\vep_3)}{4 \vep_2 \vep_3} \\
\end{array}
\right)^T
v_{ni}=\left(
\begin{array}{ccc}
 0 & 0 & 0 \\
 0 & -2 d_1 & -2 d_1-2 d_2 \\
 0 & 0 & -2 d_1 \\
 0 & -2 d_1 & -2 d_2 \\
\end{array}
\right)^T
\eeq

\beq
    g_{ni}^{(3)}=\left(
\begin{array}{ccc}
 \frac{(\vep_1+\vep_2) (\vep_2+\vep_3)}{4 \vep_1 \vep_2} & \frac{\vep_2+\vep_3}{2 \vep_2} & 1 \\
 \frac{(\vep_2-\vep_1) (\vep_2+\vep_3)}{4 \vep_1 \vep_2} & \frac{\vep_3-\vep_2}{2 \vep_2} & 1 \\
 \frac{(\vep_1+\vep_2) (\vep_3-\vep_2)}{4 \vep_1 \vep_2} & \frac{\vep_2+\vep_3}{2 \vep_2} & 1 \\
 \frac{(\vep_2-\vep_1) (\vep_3-\vep_2)}{4 \vep_1 \vep_2} & \frac{\vep_3-\vep_2}{2 \vep_2} & 1 \\
\end{array}
\right)^T
\quad
w_{ni}^{(3)}=
\left(
\begin{array}{ccc}
 0 & 0 & 0 \\
 -2 d_2-2 d_3 & -2 d_3 & 0 \\
 -2 d_3 & 0 & 0 \\
 -2 d_2 & -2 d_3 & 0 \\
\end{array}
\right)^T
\eeq
\bea
   D(k)  &= & \frac{(\vep_1+\vep_2) (\vep_2-\vep_3) \sinh (k(-d_1-d_2+d_3) )}{4 \vep_2 \vep_3}+\frac{(\vep_1-\vep_2) (\vep_2-\vep_3) \sinh
   (k(d_1-d_2+d_3) )}{4 \vep_2 \vep_3}
   \\
   & + & \frac{(\vep_1-\vep_2) (\vep_2+\vep_3) \sinh (k(-d_1+d_2+d_3) )}{4 \vep_2 \vep_3}+\frac{(\vep_1+\vep_2) (\vep_2+\vep_3)
   \sinh (k(d_1+d_2+d_3) )}{4 \vep_2 \vep_3}  \\
  f_{11}(k, z) & = & 
  \frac{
  \sinh (k
   (z-\text{z_0}))\left(\frac{(\vep_3-\vep_2) \sinh (k(d_2-d_3) )}{2 \vep_2}+\frac{(\vep_2+\vep_3) \sinh (k(d_2+d_3) )}{2 \vep_2}\right) 
   }{\vep_3D(k)} 
   \\
   f_{12}(k, z)  & = &    
   \frac{
   \sinh (d_1 k) \left(\frac{(\vep_3-\vep_2) \sinh (k (d_1+d_2-d_3-z+\text{z_0}))}{2
   \vep_2}+\frac{(\vep_2+\vep_3) \sinh (k (d_1+d_2+d_3-z+\text{z_0}))}{2 \vep_2}\right)
   }{\vep_3D(k)} 
   \\
   f_{13}(k, z)  & = &   
   \frac{
   \sinh (d_1 k)
   \sinh (k (d_1+d_2+d_3-z+\text{z_0}))
   }{\vep_3D(k)}
   \\
    f_{21}(k, z) & = &
    \frac{ 
    \sinh (d_3 k) \sinh (k (z-\text{z_0})) 
    }{\vep_3D(k)}
   \\
    f_{22}(k, z) & = & 
    \frac{ 
    \sinh (d_3 k) \left(\frac{(\vep_1-\vep_2) \sinh (k (-2 d_1+z-\text{z_0}))}{2 \vep_2}+\frac{(\vep_1+\vep_2)
   \sinh (k (z-\text{z_0}))}{2 \vep_2}\right)
    }{\vep_3D(k)}
   \\
   f_{23}(k, z) & = & 
    \frac{ 
     \sinh (k (d_1+d_2+d_3-z+\text{z_0}))\left(\frac{(\vep_1-\vep_2) \sinh (k(d_2-d_1) )}{2 \vep_2}+\frac{(\vep_1+\vep_2) \sinh (k(d_1+d_2) )}{2
   \vep_2}\right)
   }{\vep_3D(k)} \\
h_1(k)&=&{\varepsilon_1\over\varepsilon_3D(k)}
\Bigg[
{\varepsilon_2+\varepsilon_3\over 2\varepsilon_2}
\sinh\left(k(d_2+d_3)\right)
 + {\varepsilon_3-\varepsilon_2\over 2\varepsilon_2}
\sinh\left(k(d_2-d_3)\right)
\Bigg] \\
h_2(k)&=&
{\varepsilon_1\over\varepsilon_3D(k)}
\sinh\left(kd_3\right)
\eea   
For $N=4$, we show only the coefficient matrices
\beq
   c_{ni} =\left(
\begin{array}{cccc}
 1 & \frac{\vep_1+\vep_2}{2 \vep_2} & \frac{(\vep_1+\vep_2) (\vep_2+\vep_3)}{4 \vep_2 \vep_3} & \frac{(\vep_1+\vep_2) (\vep_2+\vep_3) (\vep_3+\vep_4)}{8 \vep_2 \vep_3 \vep_4} \\
 1 & \frac{\vep_1-\vep_2}{2 \vep_2} & \frac{(\vep_1+\vep_2) (\vep_2-\vep_3)}{4 \vep_2 \vep_3} & \frac{(\vep_1+\vep_2) (\vep_2+\vep_3) (\vep_3-\vep_4)}{8 \vep_2 \vep_3 \vep_4} \\
 1 & \frac{\vep_1+\vep_2}{2 \vep_2} & \frac{(\vep_1-\vep_2) (\vep_2+\vep_3)}{4 \vep_2 \vep_3} & \frac{(\vep_1+\vep_2) (\vep_2-\vep_3) (\vep_3+\vep_4)}{8 \vep_2 \vep_3 \vep_4} \\
 1 & \frac{\vep_1-\vep_2}{2 \vep_2} & \frac{(\vep_1-\vep_2) (\vep_2-\vep_3)}{4 \vep_2 \vep_3} & \frac{(\vep_1+\vep_2) (\vep_2-\vep_3) (\vep_3-\vep_4)}{8 \vep_2 \vep_3 \vep_4} \\
 1 & \frac{\vep_1+\vep_2}{2 \vep_2} & \frac{(\vep_1+\vep_2) (\vep_2+\vep_3)}{4 \vep_2 \vep_3} & \frac{(\vep_1-\vep_2) (\vep_2+\vep_3) (\vep_3+\vep_4)}{8 \vep_2 \vep_3 \vep_4} \\
 1 & \frac{\vep_1-\vep_2}{2 \vep_2} & \frac{(\vep_1+\vep_2) (\vep_2-\vep_3)}{4 \vep_2 \vep_3} & \frac{(\vep_1-\vep_2) (\vep_2+\vep_3) (\vep_3-\vep_4)}{8 \vep_2 \vep_3 \vep_4} \\
 1 & \frac{\vep_1+\vep_2}{2 \vep_2} & \frac{(\vep_1-\vep_2) (\vep_2+\vep_3)}{4 \vep_2 \vep_3} & \frac{(\vep_1-\vep_2) (\vep_2-\vep_3) (\vep_3+\vep_4)}{8 \vep_2 \vep_3 \vep_4} \\
 1 & \frac{\vep_1-\vep_2}{2 \vep_2} & \frac{(\vep_1-\vep_2) (\vep_2-\vep_3)}{4 \vep_2 \vep_3} & \frac{(\vep_1-\vep_2) (\vep_2-\vep_3) (\vep_3-\vep_4)}{8 \vep_2 \vep_3 \vep_4} \\
\end{array}
\right)^T
\eeq
\beq
  v_{ni} =\left(
\begin{array}{cccc}
 0 & 0 & 0 & 0 \\
 0 & -2 d_1 & -2 d_1-2 d_2 & -2 d_1-2 d_2-2 d_3 \\
 0 & 0 & -2 d_1 & -2 d_1-2 d_2 \\
 0 & -2 d_1 & -2 d_2 & -2 d_3 \\
 0 & 0 & 0 & -2 d_1 \\
 0 & -2 d_1 & -2 d_1-2 d_2 & -2 d_2-2 d_3 \\
 0 & 0 & -2 d_1 & -2 d_2 \\
 0 & -2 d_1 & -2 d_2 & -2 d_1-2 d_3 \\
\end{array}
\right)^T
\eeq
\beq
  g_{ni}^{(4)} =\left(
\begin{array}{cccc}
 \frac{(\vep_1+\vep_2) (\vep_2+\vep_3) (\vep_3+\vep_4)}{8 \vep_1 \vep_2 \vep_3} & \frac{(\vep_2+\vep_3) (\vep_3+\vep_4)}{4 \vep_2 \vep_3} & \frac{\vep_3+\vep_4}{2 \vep_3} & 1 \\
 \frac{(\vep_2-\vep_1) (\vep_2+\vep_3) (\vep_3+\vep_4)}{8 \vep_1 \vep_2 \vep_3} & \frac{(\vep_3-\vep_2) (\vep_3+\vep_4)}{4 \vep_2 \vep_3} & \frac{\vep_4-\vep_3}{2 \vep_3} & 1 \\
 \frac{(\vep_1+\vep_2) (\vep_3-\vep_2) (\vep_3+\vep_4)}{8 \vep_1 \vep_2 \vep_3} & \frac{(\vep_2+\vep_3) (\vep_4-\vep_3)}{4 \vep_2 \vep_3} & \frac{\vep_3+\vep_4}{2 \vep_3} & 1 \\
 \frac{(\vep_2-\vep_1) (\vep_3-\vep_2) (\vep_3+\vep_4)}{8 \vep_1 \vep_2 \vep_3} & \frac{(\vep_3-\vep_2) (\vep_4-\vep_3)}{4 \vep_2 \vep_3} & \frac{\vep_4-\vep_3}{2 \vep_3} & 1 \\
 \frac{(\vep_1+\vep_2) (\vep_2+\vep_3) (\vep_4-\vep_3)}{8 \vep_1 \vep_2 \vep_3} & \frac{(\vep_2+\vep_3) (\vep_3+\vep_4)}{4 \vep_2 \vep_3} & \frac{\vep_3+\vep_4}{2 \vep_3} & 1 \\
 \frac{(\vep_2-\vep_1) (\vep_2+\vep_3) (\vep_4-\vep_3)}{8 \vep_1 \vep_2 \vep_3} & \frac{(\vep_3-\vep_2) (\vep_3+\vep_4)}{4 \vep_2 \vep_3} & \frac{\vep_4-\vep_3}{2 \vep_3} & 1 \\
 \frac{(\vep_1+\vep_2) (\vep_3-\vep_2) (\vep_4-\vep_3)}{8 \vep_1 \vep_2 \vep_3} & \frac{(\vep_2+\vep_3) (\vep_4-\vep_3)}{4 \vep_2 \vep_3} & \frac{\vep_3+\vep_4}{2 \vep_3} & 1 \\
 \frac{(\vep_2-\vep_1) (\vep_3-\vep_2) (\vep_4-\vep_3)}{8 \vep_1 \vep_2 \vep_3} & \frac{(\vep_3-\vep_2) (\vep_4-\vep_3)}{4 \vep_2 \vep_3} & \frac{\vep_4-\vep_3}{2 \vep_3} & 1 \\
\end{array}
\right)^T
\eeq
\beq
   w_{ni}^{(4)} =\left(
\begin{array}{cccc}
 0 & 0 & 0 & 0 \\
 -2 d_2-2 d_3-2 d_4 & -2 d_3-2 d_4 & -2 d_4 & 0 \\
 -2 d_3-2 d_4 & -2 d_4 & 0 & 0 \\
 -2 d_2 & -2 d_3 & -2 d_4 & 0 \\
 -2 d_4 & 0 & 0 & 0 \\
 -2 d_2-2 d_3 & -2 d_3-2 d_4 & -2 d_4 & 0 \\
 -2 d_3 & -2 d_4 & 0 & 0 \\
 -2 d_2-2 d_4 & -2 d_3 & -2 d_4 & 0 \\
\end{array}
\right)^T
\eeq

\end{document}